\documentclass[runningheads]{llncs}
\usepackage{graphicx}
\usepackage[utf8]{inputenc}
\usepackage[T1]{fontenc}
\usepackage{booktabs}
\usepackage[labelfont=bf]{caption}
\usepackage{amsmath,amssymb,amsfonts}
\usepackage{algorithmic}
\usepackage{graphicx}
\usepackage{multirow}
\usepackage{epstopdf}
\usepackage[overload]{textcase}
\usepackage{subcaption}
\usepackage{textcomp}
\usepackage{placeins}
\usepackage{lineno,hyperref}
\usepackage{xcolor}
\usepackage{comment}
\usepackage[numbers,sort&compress]{natbib}
\usepackage{tablefootnote}
\makeatletter
\newcommand{\printfnsymbol}[1]{%
  \textsuperscript{\@fnsymbol{#1}}%
}
\makeatother

\begin{document}
\title{HI-Net: Hyperdense Inception $3D$ UNet for Brain Tumor Segmentation}
%
%
\author{Saqib Qamar\inst{1}\orcidID{0000-0002-5980-5976}\thanks{equal contribution}
\and
Parvez Ahmad\inst{2}\orcidID{0000-0003-1409-3175}\printfnsymbol{1}
\and
Linlin Shen\inst{1}}

\authorrunning{Saqib Qamar et al.}
%
%

\institute{Computer Vision Institute,
\\ School of Computer Science and Software Engineering,
\\ Shenzhen University
 \\ {sqbqamar@hust.edu.cn, llshen@szu.edu.cn}
\and
National Engineering Research Center for Big Data Technology and System,
\\ Services Computing Technology and System Lab,
\\ Cluster and Grid Computing Lab,
\\ School of Computer Science and Technology,
\\ Huazhong University of Science and Technology,
\\ 430074, Wuhan, China
\\
{parvezamu@hust.edu.cn}}
\maketitle              
\begin{abstract}
The brain tumor segmentation task aims to classify tissue into the whole tumor (WT), tumor core (TC) and enhancing tumor (ET) classes using multimodel MRI images. Quantitative analysis of brain tumors is critical for clinical decision making. While manual segmentation is tedious, time-consuming, and subjective, this task is at the same time very challenging to automatic segmentation methods. Thanks to the powerful learning ability, convolutional neural networks (CNNs), mainly fully convolutional networks, have shown promising brain tumor segmentation. This paper further boosts the performance of brain tumor segmentation by proposing hyperdense inception $3D$ UNet (HI-Net), which captures multi-scale information by stacking factorization of $3$D weighted convolutional layers in the residual inception block. We use hyper dense connections among factorized convolutional layers to extract more contexual information, with the help of features reusability. We use a dice loss function to cope with class imbalances. We validate the proposed architecture on the multi-modal brain tumor segmentation challenges (BRATS) $2020$ testing dataset. Preliminary results on the BRATS $2020$ testing set show that achieved by our proposed approach, the dice (DSC) scores of ET, WT, and TC are $0.79457$, $0.87494$, and $0.83712$, respectively.
\keywords{Brain tumor \and $3$D UNet\and dense connections \and Factorized convolutional \and Deep learning}
\end{abstract}
\section{Introduction}
Primary and secondary are two types of brain tumors. Primary brain tumors originate from brain cells, whereas secondary tumors metastasize into the brain from other organs. Gliomas are primary brain tumors. Gliomas can be further sub-divided into two parts: low-grade (LGG) and high-grade (HGG). High-grade gliomas are an aggressive type of malignant brain tumor that proliferates, usually requires surgery and radiotherapy, and has a poor survival prognosis. Magnetic resonance imaging (MRI) is a critical diagnostic tool for brain tumor analysis, monitoring, and surgery planning. Usually, several complimentary $3D$ MRI modalities - such as $T1$, $T1$ with contrast agent $(T1c)$, $T2$, and fluid attenuation inversion recover $(FLAIR)$ are required to emphasize different tissue properties and areas of tumor spread. For example, the contrast agent, usually gadolinium, emphasizes hyperactive tumor subregions in $T1c$ MRI modality.

Deep learning techniques, especially CNNs, are prevalent for the automatic segmentation of brain tumors. CNN can learn from examples and demonstrate state-of-the-art segmentation accuracy both in $2$D natural images and in $3$D medical image modalities. The information of segmentation provides an accurate, reproducible solution for further tumor analysis and monitoring. Multi-modal brain tumor segmentation challenge (BRATS) aims to evaluate state-of-the-art methods for the segmentation of brain tumors by providing a $3$D MRI dataset with ground truth labels annotated by physicians \cite{bakas2017_segmentation}, \cite{bakas2017segmentation}, \cite{Bakas2017}, \cite{DBLP:journals/corr/abs-1811-02629}, \cite{6975210}.  A $3D$ UNet is a popular CNN architecture for automatic brain tumor segmentation \cite{DBLP:journals/corr/abs-1802-10508}. The multi-scale contextual information of the encoder-decoder sub-networks is effective for the accurate brain tumor segmentation task. Several variations of the encoder-decoder architectures were proposed for MICCAI BraTS $2018$ and $2019$ competitions.The potential of several deep architectures \cite{kamnitsas2017efficient,DBLP:journals/corr/LongSD14,DBLP:journals/corr/RonnebergerFB15} and their ensembling procedures for brain tumor segmentation was discussed by a top-performing method \cite{DBLP:journals/corr/abs-1711-01468} for MICCAI BRATS $2017$ competition. Wang \textit{et al}\cite{DBLP:journals/corr/abs-1709-00382} proposed architectures with factorized weighted layers to save the GPU memory and the computational time. At the same time, the majority of these architectures used either the bigger input sizes \cite{DBLP:journals/corr/abs-1810-11654} or cascaded training \cite{10.1007/978-3-030-46640-4_22} or novel pre-processing \cite{10.1007/978-3-030-11726-9_25} and post-processing strategies \cite{10.1007/978-3-030-11726-9_21} to improve the segmentation accuracy. In contrast, few architectures demonstrate the important memory consumption of $3$D convolutional layers. Chen \textit{et al} \cite{10.1007/978-3-030-11726-9_32} used an important concept in which each weighted layer was split into three branches in a parallel fashion, each with a different orthogonal view, namely axial, sagittal, and coronal. However, more complex combinations exist between features within and in-between different orthogonal views, which can significantly increase the learning representation \cite{DBLP:journals/corr/abs-1804-02967}. Inspired by the S$3$D UNet architecture \cite{10.1007/978-3-030-11726-9_32,DBLP:journals/corr/abs-1712-04851}, we propose a variant encoder-decoder based architecture for the brain tumor segmentation. The key contributions of our study are as follows:
\begin{itemize}
\item A novel hyperdense inception $3$D UNet (HI-Net) architecture is proposed by stacking factorization of $3$D weighted convolutional layers in the residual inception block.
\item In each residual inception block, hyper-dense connections are used in-between different orthogonal views to learn more complex feature representation.
\item Our network achieves state-of-the-art performance as compared to other recent methods.
\end{itemize}

\section{Proposed Method}
\begin{figure*}[!t]
\includegraphics[width=\textwidth]{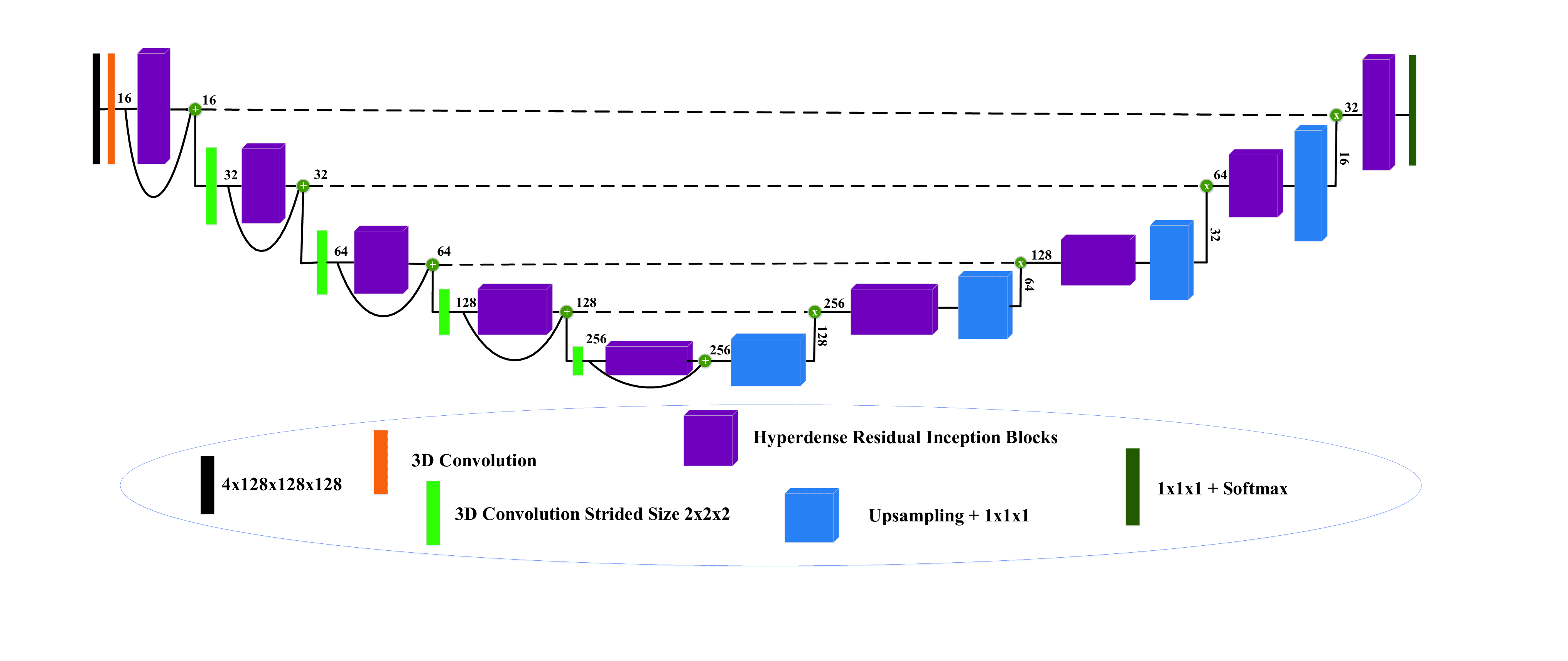}
\caption{Proposed HI-Net architecture. The element-wise addition operations ($+$ symbol with the oval shape) are employed to design the proposed architecture. The modified residual inception blocks  (violet), known as hyperdense residual inception blocks, are used in the encoder-decoder paths. The length of the encoder path is longer than the decoder part by performing repetitions on several levels. The maximum repetition is $4$ on the last level of the encoder part to draw the semantic information from the lowest input resolution. Finally, the softmax activation is performed for the outcomes.} \label{fig1}
\end{figure*}

\begin{figure*}[!htbp]
\includegraphics[clip, trim=5cm 0cm 0cm 0cm, width=1.3\textwidth]{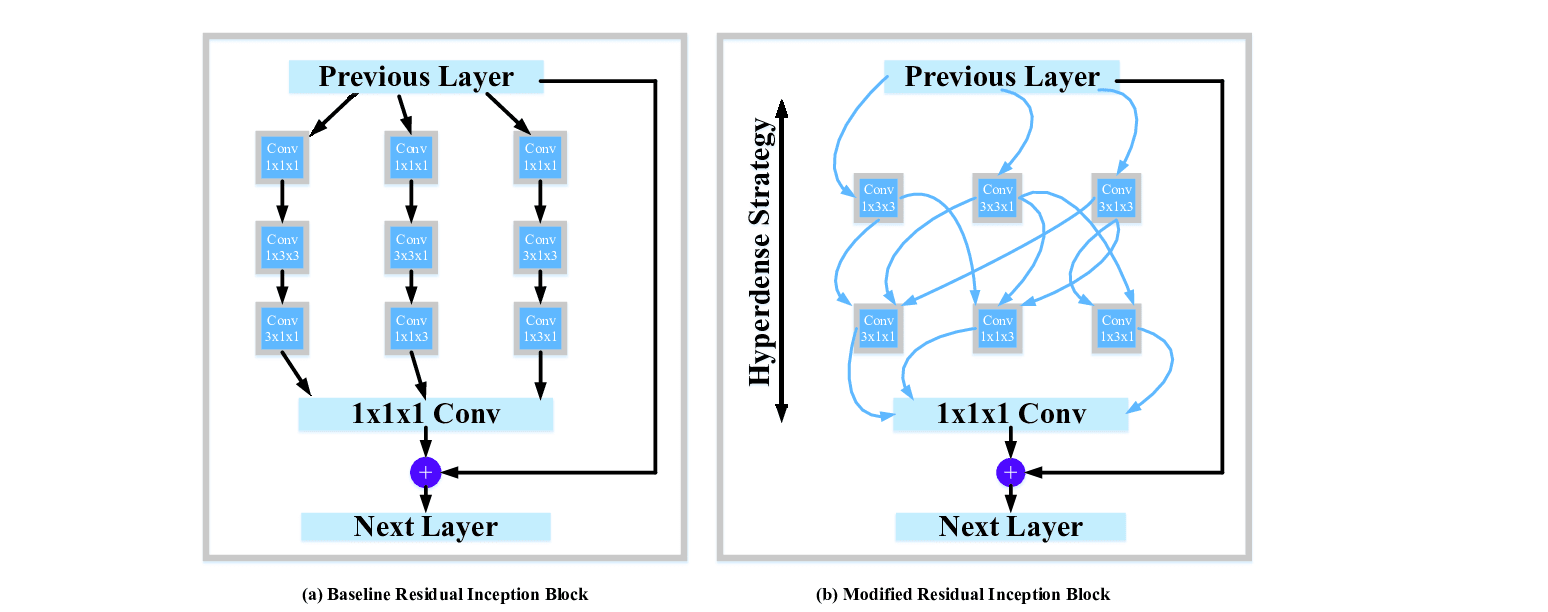}
\caption{Difference between baseline and modified residual inception blocks.  ($a$) represent a baseline residual inception block with a separable $3$D convolutional layer, while the proposed block with inter-connected dense connections is shown in ($b$).} \label{fig2}
\end{figure*}

Figure \ref{fig1} shows the proposed HI-Net architecture for brain tumor segmentation. The network's left side works as an encoder to extract the features of different levels, and the right component of the network acts as a decoder to aggregate the features and the segmentation mask. The modified residual inception blocks of the encoder-decoder sub-networks have two $3D$ convolutional layers, and each layer has followed the structure of Fig. \ref{fig2}(b). In contrast, traditional residual inception blocks are shown in Fig. \ref{fig2}(a). This study employed inter-connections of dense connections within and in-between different orthogonal views to learn more complex feature representation. In the stage of encoding, the encoder extracts feature at multiple scales and create fine-to-coarse feature maps. Fine feature maps contain low-level features but more spatial information, while coarse feature maps provide the opposite. Skip connection is used to combine coarse and fine feature maps for accurate segmentation. Unlike standard residual UNet, the encoder sub-network uses a self-repetition procedure on multiple levels to generate semantic maps for fine feature maps and thus select relevant regions in the fine feature maps to concatenate with the coarse feature maps.

\section{Implementation Details}
\subsection{Dataset}

The BRATS 2020 \cite{bakas2017_segmentation}, \cite{bakas2017segmentation}, \cite{Bakas2017}, \cite{DBLP:journals/corr/abs-1811-02629}, \cite{6975210} training dataset included $369$ cases ($293$ HGG and $76$ LGG), each with four rigidly aligned $3$D MRI modalities ($T1$, $T1c$, $T2$, and $FLAIR$), resampled to $1\times1\times1$ mm isotropic resolution and skull-stripped. The input image size is $240\times240\times155$. The data were collected from $19$ institutions, using various MRI scanners. Annotations include $3$ tumor subregions: WT, TC, and ET. Two additional datasets without the ground truth labels are provided for validation and testing. These datasets required participants to upload the segmentation masks to the organizers' server for evaluations.  In validation ($125$ cases) and testing ($166$) datasets, each subject includes the same four modalities of brain MRI scans but no ground truth. In our experiment, the training set is applied to optimize the trainable parameters in the network. The validation and testing sets are utilized to evaluate the performance of the trained network.
\subsection{Experiments}
The network is implemented by Keras and trained on Tesla V$100$–SXM$2$ $32$ GB GPU card with a batch size of $1$. Adam optimizer with an initial learning rate $3\times10{^{-5}}$  is employed to optimize the parameters. The learning rate is reduced by $0.5$ per $30$ epochs. The network is trained for $350$ epochs. During network training, augmentation techniques such as random rotations and mirroring are employed. The size of the input during the training of the network is $128\times128\times128$. The multi-label dice loss function \cite{DBLP:journals/corr/MilletariNA16} addressed the class imbalance problem. Equation \ref{e1} shows the mathematical representation of loss function.

\begin{equation}\label{e1}
     Loss=-\frac{2}{D} \sum_{d \in D}
            \frac{\sum_{j}P_{(j, d)}T_{(j, d)} + r}{{\sum_{j}P_{(j, d)}} + {\sum_{j}T_{(j, d)}} + r}      \linebreak
\end{equation}

where $P_{(j, d)}$ and $T_{(j, d)}$ are the prediction obtained by softmax activation and ground truth at voxel $j$ for class $d$, respectively. $D$ is the total number of classes.

\subsection{Evaluation Metrics}
Multiple criteria are computed as performance metrics to quantify the segmentation result. Dice coefficient (DSC) is the most frequently used metric for evaluating medical image segmentation. It measures the overlap between the segmentation and ground truth with a value between $0$ and $1$. The higher the Dice score, the better the segmentation performance. \textit{Sensitivity} and \textit{specificity} are also commonly used statistical measures. The sensitivity called true positive rate is defined as the proportion of positives that are correctly predicted. It measures the portion of tumor regions in the ground truth that is also predicted as tumor regions by the segmentation method. The specificity, called true negative rate, is defined as the proportion of correctly predicted negatives. It measures the portion of normal tissue regions in the ground truth that is also predicted as normal tissue regions by the segmentation method.

\subsection{Results}
\begin{figure*}[!t]
\includegraphics[clip, trim=0cm 5cm 0cm 3cm, width=1\textwidth]{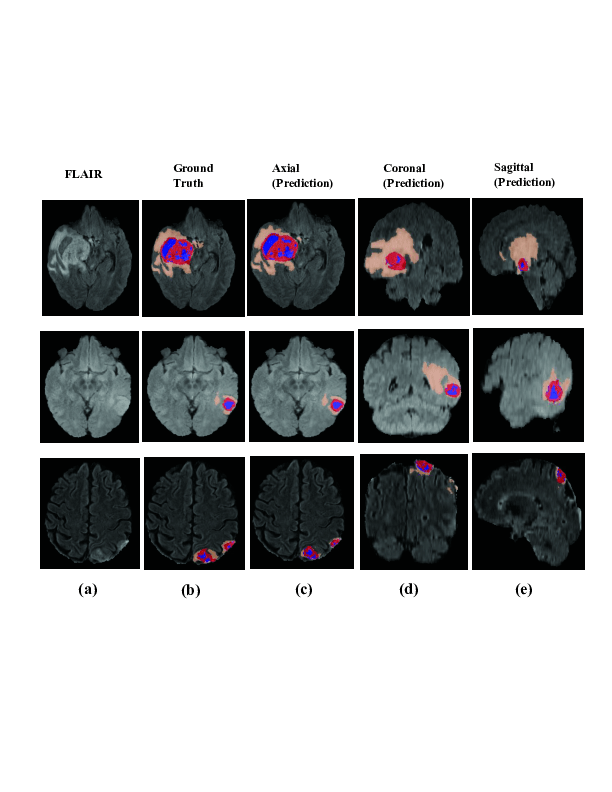}
\caption{Segmentation results on the training dataset of the BRATS $2020$. From left to right: Ground-truth and predicted results on FLAIR modality; WT (brown), TC (red) and ET (blue).} \label{fig3}
\end{figure*}

The performance of our proposed architecture is evaluated on training, validation, and the testing sets provided by BRATS $2020$. Table \ref{tab1} presents the quantitative analysis of the proposed work. We have secured mean DSC scores of ET, WT, and TC as $0.74191$, $0.90673$, and $0.84293$, respectively, on the validation dataset, while  $0.80009$, $0.92967$, and $0.90963$ on the training dataset. At the same time, our proposed approach obtained mean DSC scores of ET, WT, and TC as $0.79457$, $0.87494$, and $0.83712$, respectively, on the testing dataset.  In Table \ref{tab1}, sensitivity and specificity are also presented on training, validation, and the testing datasets. Table \ref{tab2} shows the comparable study of proposed work with the baseline work \cite{10.1007/978-3-030-11726-9_32}. Our proposed HI-Net achieves higher scores for each tumor than the baseline work. Furthermore, ablation studies are conducted to assess the modified residual inception blocks' influence with and without the inter-connected dense connections. The influence of these connections on DSCs of ET, WT, and TC is shown in Table \ref{tab2}. To provide qualitative results of our method, three-segmented images from training data are shown in Fig \ref{fig3}. In summary, modified inception blocks significantly improve the DSCs of ET, WT, and TC against the baseline inception blocks.

\setlength{\tabcolsep}{12pt}
\begin{table*}[!t]
\caption{BRATS $2020$ training, validation and testing results. Mean average scores on different metrics.}\label{tab1}
\begin{tabular}{llllll}
\hline
Dataset&{Metrics}&\multicolumn{1}{l}{WT}&
\multicolumn{1}{l}{TC}&
\multicolumn{1}{l}{ET}\\
\hline
BRATS 2020 Training&DSC     &92.967&90.963&80.009\\
& Sensitivity               &93.004&91.282&80.751\\
& Specificity               &99.932&99.960&99.977\\[6pt]
BRATS 2020 Validation&DSC   &90.673&84.293&74.191\\
& Sensitivity               &90.485&80.572&73.516\\
& Specificity               &99.929&99.974&99.977\\[6pt]
BRATS 2020 Testing&DSC   &87.494&83.712&79.457\\
& Sensitivity               &91.628&85.257&82.409\\
& Specificity               &99.883&99.962&99.965\\\hline
\end{tabular}
\end{table*}

\begin{table}[!t]
\caption{Performance evaluation of different methods on the BRATS $2020$ validation dataset. For comparison, only DSC scores are shown. All scores are evaluated online.} \label{tab2}
\centering
\resizebox{0.8\textwidth}{!}{
\begin{tabular}{lllll}
\hline
{Methods} & {ET} & {WT} & {TC} \\
\hline
Baseline Work        & 70.616 & 90.670 & 82.136\\ [2pt]
Proposed Work         &74.191&90.673 &84.293 \\
\hline
\end{tabular}}
\end{table}

\section{Conclusion}
We proposed a HI-Net architecture for brain tumor segmentation. Each $3$D convolution is splitted into three parallel branches in the residual inception block, each with different orthogonal views, namely axial, sagittal and coronal. We also proposed hyperdense connections among factorized convolutional layers to extract more contextual information. The HI-Net architecture secures high DSC scores for all types of tumors. This network has been evaluated on the BRATS $2020$ Challenge testing dataset and achieved average DSC scores of $0.79457$, $0.87494$, and $0.83712$ for the segmentation of ET, WT, and TC, respectively. Compared with the performance of the validation dataset, the scores on the testing set are higher. In the future, we will work to enhance the robustness of the network to improve the segmentation performance by using some post-processing methods such as a fully connected conditional random field (CRF).
\section*{Acknowledgment}
This work is supported by the National Natural Science Foundation of China under Grant No. 91959108.
\bibliographystyle{splncs04}
\bibliography{Nature}
\end{document}